\documentclass[runningheads]{llncs}
\usepackage{graphicx}
\usepackage{subcaption,booktabs}
\usepackage{xcolor}
\usepackage{hyperref}
\hypersetup{
    colorlinks=true,
    linkcolor=magenta,
    filecolor=magenta,      
    urlcolor=magenta,
    pdftitle={Overleaf Example},
    pdfpagemode=FullScreen,
    }
\begin{document}
\title{IndicIRSuite: Multilingual Dataset and Neural Information Models for Indian Languages}
%
%

\author{Saiful Haq\and
Ashutosh Sharma\and
Pushpak Bhattacharyya}

%
\authorrunning{Haq et al.}
%

\institute{Department of Computer Science and Engineering, IIT Bombay, India\\
\email{\{saifulhaq, pb\}@cse.iitb.ac.in, sharma96@illinois.edu}}

\maketitle              
\begin{abstract}
In this paper, we introduce Neural Information Retrieval resources for 11 widely spoken Indian Languages (Assamese, Bengali, Gujarati, Hindi, Kannada, Malayalam, Marathi, Oriya, Punjabi, Tamil, and Telugu) from two major Indian language families (Indo-Aryan and Dravidian). These resources include (a) INDIC-MARCO, a multilingual version of the MSMARCO dataset in 11 Indian Languages created using Machine Translation, and (b) Indic-ColBERT, a collection of 11 distinct Monolingual Neural Information Retrieval models, each trained on one of the 11 languages in the INDIC-MARCO dataset. To the best of our knowledge, IndicIRSuite is the first attempt at building large-scale Neural Information Retrieval resources for a large number of Indian languages, and we hope that it will help accelerate research in Neural IR for Indian Languages. Experiments demonstrate that Indic-ColBERT achieves 47.47\% improvement in the MRR@10 score averaged over the INDIC-MARCO baselines for all 11 Indian languages except Oriya, 12.26\% improvement in the NDCG@10 score averaged over the MIRACL Bengali and Hindi Language baselines, and 20\% improvement in the MRR@100 Score over the Mr.Tydi Bengali Language baseline. \linebreak IndicIRSuite is available at \href{https://github.com/saifulhaq95/IndicIRSuite}{github.com/saifulhaq95/IndicIRSuite}.
\end{abstract}
\section{Introduction}
Information Retrieval (IR) models process user queries and search the document corpus to retrieve a ranked list of relevant documents ordered by a relevance score. Classical IR models, like BM25
\cite{robertson2009probabilistic}, retrieve documents that have lexical overlap with the query tokens. Recently, there has been a notable upsurge in adopting Neural IR models utilizing language models such as BERT \cite{devlin2018bert}, which enable semantic matching of queries and documents. This paradigm shift has demonstrated substantial efficacy in retrieving and re-ranking documents. ColBERTv2\cite{santhanam2021colbertv2}, one of the state-of-art neural IR models, has shown 0.185 points improvement in Normalized Discounted Cumulative Gain@10 (NDCG@10) Score over the BM25 Model baseline on the MSMARCO passage ranking dataset \cite{thakur2021beir}. 

The importance of dataset size outweighs domain-matching in training neural IR models \cite{zhang2022best}. Due to the scarcity of large-scale domain-specific datasets, Neural IR models are first trained on the MSMARCO passage ranking dataset \cite{nguyen2016ms}, and they are subsequently evaluated on domain-specific datasets in a zero-shot manner. MSMARCO dataset contains 39 million training triplets (q, +d, -d) where q is an actual query from the Bing search engine, +d is a human-labeled passage answering the query, and -d is sampled from unlabelled passages retrieved by the BM25 Model. Notably, the MSMARCO dataset is in the English Language, meaning neural IR models trained on it supposedly work well with only English queries and passages.

Monolingual IR for non-English languages \cite{zhang2022making} \cite{zhang2021mr}, Multilingual IR \cite{lawrie2023neural}, and Cross-lingual IR \cite{lin2023simple} \cite{sun2020clirmatrix} extend the English IR paradigm to support diverse languages. In Monolingual IR for non-English languages, the query and passages are in the same language, which is not English. In Cross-lingual IR, the query is used to create a ranked list of documents such that each document is in the same language, different from the query language. In Multilingual IR, the query is used to create a ranked list of documents such that each document is in one of the several languages, which can be the same or different from the query language. In this work, we focus on Monolingual IR for non-English languages.
Monolingual IR for non-English languages involves training an encoder like mBERT \cite{devlin2018bert}, on a large-scale general-domain monolingual dataset for non-English languages to minimize the pairwise softmax cross-entropy loss. The trained models are subsequently finetuned or used in a zero-shot manner on small-scale domain-specific datasets. However, there is a notable lack of large-scale datasets like mMARCO \cite{bonifacio2021mmarco} for training monolingual neural IR models on many low-resource Indian languages. We introduce neural IR resources to address this scarcity and facilitate Monolingual neural IR across 11 Indian languages. Our contributions are:
\begin{itemize}
\item INDIC-MARCO, a multilingual dataset for training neural IR models in 11 Indian Languages (Assamese, Bengali, Gujarati, Hindi, Kannada, Malayalam, Marathi, Oriya, Punjabi, Tamil and Telugu). For every language in INDIC-MARCO, there exists 8.8 Million passages, 1 Million queries, 39 million training triplets (query, relevant document, irrelevant document), and approximately one relevant document per query. To the best of our knowledge, this is the first large-scale dataset for training a neural IR system on 11 widely spoken Indian languages.
\item Indic-ColBERT, a collection of 11 distinct Monolingual Neural Information Retrieval models, each trained on one of the 11 languages in the INDIC-MARCO dataset. Indic-ColBERT achieves 47.47\% improvement in the MRR @10 score averaged over the INDIC-MARCO baseline for all 11 Indian languages except Oriya, 12.26\% improvement in the NDCG @10 score averaged over the MIRACL Bengali and Hindi Language baselines, and 20\% improvement in the MRR@100 Score over the Mr.Tydi Bengali Language baseline. To the best of our knowledge, this is the first effort for a neural IR dataset and models on 11 widely spoken Indian languages, thereby providing a benchmark for Indian language IR. 
\end{itemize}
\section{Related work}
The size of datasets holds greater importance than ensuring domain matching in the training of neural IR models \cite{zhang2022best}. In terms of size and domain, mMARCO \cite{bonifacio2021mmarco} is the most similar to our work as it introduces a large-scale machine-translated version of MSMARCO in many languages, Hindi being the only Indian language. MIRACL \cite{zhang2022making} and Mr.Tydi \cite{zhang2021mr} also introduce datasets and models for Monolingual Neural IR in Hindi, Bengali, and Telugu.

FIRE\footnote{http://fire.irsi.res.in/fire/static/data} was the most active initiative from 2008 to 2012 for Multilingual IR in Indian languages. FIRE developed datasets for Multilingual IR in six Indian Languages (Bengali, Gujarati, Hindi, Marathi, Oriya, and Tamil). However, the size of these datasets is not large enough to train neural IR systems based on transformer models like mBERT\cite{devlin2018bert} and XLM\cite{lample2019cross}. In addition, the text in the FIRE dataset comes from newspaper articles \cite{palchowdhury2013overview}, which is domain-specific; hence, the models trained on such datasets cannot generalize well to other domains. Due to the lack of large-scale datasets, Cross-lingual knowledge transfer via Distillation has become popular for neural IR in low-resource languages \cite{huang2023crosslingual} \cite{Huang_2023}. 

The key distinction in our work from the earlier approaches is that we introduce monolingual datasets and neural IR models in 11 widely spoken Indian Languages (Assamese, Bengali, Gujarati, Hindi, Kannada, Malayalam, Marathi, Oriya, Punjabi, Tamil and Telugu), that can also benefit Cross-lingual IR and Multilingual IR models from the cross-lingual transfer effects when trained on a large number of Indian Languages \cite{zhang2022best}.




\section{INDIC-MARCO}\label{sec:indicmarco}

We introduce the INDIC-MARCO dataset, a multilingual version of the MSMARCO dataset. We translate the queries and passages in the MSMARCO passage ranking dataset into 11 widely spoken Indian languages (Assamese, Bengali, Gujarati, Hindi, Kannada, Malayalam, Marathi, Oriya, Punjabi, Tamil and Telugu) originating from two major language families (Indo-Aryan and Dravidian). The translation process utilizes the int-8 quantized version of the NLLB-1.3B-Distilled Model \cite{costa2022no}, available at CTranslate2\footnote{https://forum.opennmt.net/t/nllb-200-with-ctranslate2/5090} \cite{klein2020opennmt}. We chose int-8 quantized version of NLLB-1.3B-Distilled Model for two reasons: (a) it has shown remarkable performance in terms of BLEU scores for many Indian languages as compared to IndicBART \cite{dabre2021indicbart} and IndicTrans \cite{ramesh2022samanantar} (b) Quantization \cite{klein2020opennmt} enables faster inference with less computing power and little or no drop in translation quality. The machine translation process employs specific hyper-parameters: a beam width of 4, a maximum decoding sequence length of 200 tokens, a batch size of 64, and a batch type equal to \lq{examples}\rq. Passages from the MSMARCO dataset are split into multiple sentences using the Moses SentenceSplitter\footnote{https://pypi.org/project/mosestokenizer/}, ensuring that each sentence serves as a translation unit in a batch of 64 sentences. In contrast, queries with an average length of 5.96 words \cite{thakur2021beir} are not sentence-split before translation. We also translate the MSMARCO Dev-Set(Small)\footnote{https://ir-datasets.com/msmarco-passage.html} containing 6,390 queries (1.1 qrels/query) to obtain INDIC-MARCO Dev-set(Small). The translation process on an Nvidia A100 GPU with 12 GB VRAM takes approximately 1584 hours for passages in MSMARCO, 55 hours for queries in MSMARCO, and 1.5 hours for queries in MSMARCO Dev-Set(Small).

Upon translation, the resulting INDIC-MARCO dataset comprises around 8.8 million passages, 530k queries, and 39 Million training triplets in 11 Indian languages. This dataset allows for training monolingual neural IR models for each language in the INDIC-MARCO dataset.
\section{Model}

\subsection{Baselines}
BM25 \cite{robertson2009probabilistic} serves as a strong baseline as it performs better than many neural IR models on domain-specific datasets with exceptions\cite{thakur2021beir}. It does not require any training. BM25 retrieves documents containing query tokens and assigns them a score for re-ranking based on the frequency of query tokens appearing in them and the document length. In this work, we use the BM25 implementation provided by Pyserini\footnote{https://github.com/castorini/pyserini} with values for parameters k1=0.82 and b=0.68 for evaluation on INDIC-MARCO Dev-Set obtained after machine translation. We use Whitespace Analyzers to tokenize queries and documents during indexing and searching for all Indian languages except Hindi, Bengali, and Telugu, for which we use language-specific analyzers provided in Pyserini. BM25-tuned (BM25-T) presented in Mr.Tydi\cite{zhang2021mr} is optimized to maximize the MRR@100 score on the Mr.Tydi test-set using a grid search over the range [0.1, 0.6] for k1 and [0.1, 1] for b.

Multilingual Dense Passage Retriever (mDPR) is presented in both Mr.Tydi and MIRACL by replacing the BERT encoder in Dense Passage Retriever(DPR) \cite{karpukhin2020dense} with an mBERT encoder. In Mr.Tydi, mDPR is trained on English QA dataset \cite{kwiatkowski-etal-2019-natural} and used in a zero-shot manner for indexing and retrieval of documents. In MIRACL, mDPR is trained on the MSMARCO dataset and used in a zero-shot manner for indexing and retrieving documents. Multilingual ColBERT (mCol) is introduced in MIRACL by replacing the BERT encoder in ColBERT\cite{santhanam2021colbertv2} with an mBERT encoder. mCol is trained on the MSMARCO dataset and used in a zero-shot manner for indexing and retrieval of documents.

\begin{table}[t]
\caption{Results on INDIC-MARCO Dev-Set(Small). mColBERT (mCol) is trained on MSMARCO passage ranking dataset \cite{nguyen2016ms}. Indic-ColBERT(iCol) is a set of 11 distinct monolingual neural IR models trained on INDIC-MARCO.}\label{tab:indicmarco}
\centering
\begin{tabular}{lllllll}
\hline
Language &
\multicolumn{3}{c}{MRR@10} & \multicolumn{3}{c}{Recall@1000}\\
& BM25 & mCol & \textbf{iCol} & BM25 & mCol & \textbf{iCol}\\
\hline
Assamese & 0.078 & 0.095 & \textbf{0.176} & 0.449 &0.503 & \textbf{0.698}\\
Bengali & 0.112 & 0.159 & \textbf{0.221} & 0.622 &0.691 & \textbf{0.788}\\
Gujarati & 0.100 & 0.141 & \textbf{0.232} & 0.539 &0.653 & \textbf{0.805}\\
Hindi & 0.125 & 0.171 & \textbf{0.223} & 0.678 &0.729 & \textbf{0.772}\\
Kannada & 0.089 & 0.156 & \textbf{0.219} & 0.520 &0.691 & \textbf{0.787} \\
Malayalam & 0.076 & 0.124 & \textbf{0.198} & 0.442 &0.603 & \textbf{0.742}\\
Marathi & 0.085 & 0.143 & \textbf{0.207} & 0.476 &0.655 & \textbf{0.750}\\
Oriya & \textbf{0.086} & 0.002 & 0.002 & \textbf{0.484} &0.022 & 0.016\\
Punjabi & 0.113 & 0.134 & \textbf{0.211} & 0.603 &0.637 & \textbf{0.766}\\
Tamil & 0.088 & 0.144 & \textbf{0.202} & 0.495 &0.661 & \textbf{0.756}\\
Telugu & 0.1007 & 0.144 & \textbf{0.206} & 0.569 &0.648 & \textbf{0.749}\\
\hline
\end{tabular}
\end{table}
\subsection{Indic-ColBERT}
Indic-ColBERT (iCol) is based on ColBERTv2 \cite{santhanam2021colbertv2} with distinctions: it uses mBERT as query-document encoder, and is trained on INDIC-MARCO. Model architecture comprises (a) a query encoder, (b) a document encoder, and (c) max-sim function (same as ColBERTv2). Given a query with $q$ tokens and a document with $d$ tokens, the Query encoder outputs $q$ fix-sized token embeddings, and the document encoder outputs $d$ fix-sized token embeddings. The maximum input sequence length for the query, $q_{max}$, and, for the document, $d_{max}$, is set before giving them to the respective encoders. If $q$ is less than $q_{max}$, we append $q_{max}-q$ [MASK] tokens to the input query, and if $q$ is greater than $q_{max}$, $q$ is truncated to $q_{max}$. If $d$ is less than $d_{max}$, then $d$ is neither truncated nor padded. If $d$ is greater than $d_{max}$, $d$ is truncated to $d_{max}$. The max-sim function is used to obtain the relevance score of a document for a query using the encoded representations.
\section{Experiment Setup}
We train 11 distinct Indic-ColBERT (iCol) models separately for 50k iterations with a batch size of 128 on the first 6.4 million training triplets from the INDIC-MARCO dataset to optimize the pairwise softmax cross entropy loss function, where each triplet contains a query, a relevant passage and an irrelevant passage in one of the 11 languages on which the model is trained. The mBERT encoder is finetuned from the official "bert-base-multilingual-uncased" checkpoint, and the remaining parameters are trained from scratch.
\section{Results}
\begin{table}[t]
\caption{Results on Mr.Tydi test-set: We use official BM25, BM25-tuned(BM25-T), and mDPR model scores from Mr.Tydi \cite{zhang2021mr}. mColBERT (mCol), which is trained on MSMARCO, and Indic-ColBERT(iCol) are used in a zero-shot manner on Mr.Tydi test-set.}\label{tab:mrtydi}
\centering
\begin{tabular}{llllll|lllll}
\hline
Language &
\multicolumn{5}{c}{MRR@100} & \multicolumn{5}{c}{Recall@100}\\
& BM25 &BM25-T &mDPR &mCol &\textbf{iCol} & BM25 &BM25-T &mDPR &mCol & \textbf{iCol}\\
\hline
Bengali &  0.418 & 0.413 & 0.258 & 0.414 & \textbf{0.501} &  0.869 &  0.874 &  0.671 & 0.846 & \textbf{0.864}\\
Telugu & 0.343 & \textbf{0.424} &  0.106 & 0.314 & 0.393 &  0.758 & \textbf{0.813} &  0.352 & 0.589 & 0.688\\
\hline
\end{tabular}
\end{table}
\begin{table}[t]
\caption{Results on MIRACL dev-set: We use official scores of BM25, mDPR, and mColBERT (mCOl) models from MIRACL \cite{zhang2022making}. Indic-ColBERT(iCol) is used in a zero-shot manner on MIRACL dev-set.}\label{tab:miracl}
\centering
\begin{tabular}{lllll|llll}
\hline
Language &
\multicolumn{4}{c}{NDCG@10} & \multicolumn{4}{c}{Recall@100}\\
& BM25 &mDPR &mCol & \textbf{iCol} & BM25 &mDPR &\textbf{mCol} &iCol\\
\hline
Bengali & 0.508 & 0.443 & 0.546 & \textbf{0.606} & 0.909 & 0.819 & \textbf{0.913} & 0.894\\
Hindi & 0.458 & 0.383 & 0.470 & \textbf{0.483} & 0.868 & 0.776 & \textbf{0.884} & 0.811\\
Telugu & \textbf{0.494} & 0.356 & 0.462 & 0.479 & \textbf{0.831} & 0.762 & 0.830 & 0.768\\
\hline
\end{tabular}
\end{table}
Indic-ColBERT (iCol) outperforms baseline models (BM25, mCol) by 47.47\% in terms of MRR@10 Score on INDIC-MARCO Dev-Set(Small) (Refer Table \ref{tab:indicmarco}) averaged over all 11 Indian languages (excluding Oriya). We do not see any improvements for Oriya because mBERT used in Indic-ColBERT is not pre-trained on Oriya and Assamese. Assamese demonstrates a 125\% MRR@10 improvement over the BM25 baseline, attributed to its linguistic similarity with Bengali (indicated by the mColBERT model outperforming BM25 by 21\% in MRR@10 Score) and the high-quality data in INDIC-MARCO, further enhancing the MRR@10 score by 104\%, making INDIC-MARCO a significant contributor to the advancement for a low-resource language like Assamese which mBERT does not support.

Indic-ColBERT (iCol) outperforms baseline models (BM25, BM25-T, mDPR, mCol) by 20\%, in terms of MRR@100 Score on Mr. Tydi test-set(Refer Table \ref{tab:mrtydi}) for Bengali Language. For Telugu, Indic-ColBERT (iCol) outperforms 3 (BM25, mDPR, mCol) out of 4 baselines in terms of MRR@100 scores. 


Indic-ColBERT (iCol) outperforms baseline models (BM25, mDPR, mCol) by 19.29\% in Bengali and 5.4\% in Hindi, in terms of NDCG@10 Score on MIRACL dev-set(Refer Table \ref{tab:miracl}). For Telugu, Indic-ColBERT (iCol) outperforms 2 (mDPR, mCol) out of 3 baselines in terms of NDCG@10 scores. 

\section{Summary, conclusion, and future work}
In this paper, we introduce IndicIRSuite, comprising INDIC-MARCO, a multilingual dataset for neural IR in 11 Indian languages, and Indic-ColBERT, a set of 11 distinct monolingual neural IR models based on ColBERTv2. Our results demonstrate performance improvements over baselines in Mr.Tydi, MIRACL, and INDIC-MARCO datasets. We also showcase the utility of INDIC-MARCO for low-resource Indian languages, such as Assamese, which are not supported by multilingual models like mBERT\cite{devlin2018bert} but are linguistically similar to languages like Bengali. Future work includes extending IndicIRSuite for Cross-lingual IR, Multilingual IR, and Multilingual Question-Answering.

\bibliographystyle{acm}
\bibliography{mybibliography}
\end{document}